# Nature of the Breakdown in the Stokes-Einstein Relationship in a Hard Sphere Fluid

**Sanat K. Kumar**

Department of Chemical and Biological Engineering
Rensselaer Polytechnic Institute
Troy, NY

**Grzegorz Szamel**

Department of Chemistry
Colorado State University
Fort Collins, CO

**Jack F. Douglas**

Polymers Division
National Institutes of Standards and Technology
Gaithersburg, MD

## Abstract

Molecular Dynamics simulations of high density hard sphere fluids clearly show a breakdown of the Stokes-Einstein equation (SE). This result has been conjectured to be due to the presence of mobile particles, i.e., ones which have the propensity to "hop" distances which are integer multiples of the interparticle distance. We conclusively show that, even though the whole liquid violates the SE equation at high densities, the sedentary particles, i.e., ones complementary to the "hoppers", obey the SE relationship. These results strongly support the notion that the unusual dynamics of fluids near vitrification are caused exclusively by the presence of hopping particles.

Typical liquids, far above their glass transition, obey the Stokes-Einstein (SE) equation, D$\tau$ =constant, relating the product of the self-diffusion coefficient, D, and the structural relaxation time, $\tau$, for a homogeneous fluid at thermal equilibrium. However, recent experiments and simulations have shown that this relationship "breaks down" in a dramatic fashion in the vicinity of the glass transition temperature.[1-8] It is now commonly accepted that this phenomenon is a consequence of "dynamic heterogeneity" in supercooled liquids, specifically the presence of particles having excessively high and low mobility relative to the average motion..[9, 10] There is much current research directed at quantifying the dynamic heterogeneity phenomenon and at understanding its physical origin.[1, 3, 5-26]

To better understand the origins and the nature of the dynamic heterogeneity we utilize molecular dynamics (MD) simulations to study a model polydisperse hard sphere fluid. Our system is the same as that used in the recent work of Doliwa and Heuer.[27] All spheres were of identical mass, but the sphere diameters were sampled from a Guassian distribution with a mean of $\sigma$ and a standard deviation of 0.1$\sigma$; particle diameters were only allowed to range ± 3 standard deviations about the mean value. A variety of packing fractions, defined as $\eta = \dfrac{\sum (\pi/6)\sigma_i^3}{V}$ ranging from 0.35 to 0.60 were examined using an event driven MD algorithm.[28] Most of our simulations utilized 864 spheres, but we also performed some simulations with 2048 particles. The initial state of the system was generated by placing the spheres with randomly generated diameters on an FCC lattice so that there was no overlap. This typically resulted in a density of $\eta$ ~ 0.45 - 0.5. The systems were then simulated using MD and periodically compressed till the desired $\eta$ was attained. An important issue is the length of the MD simulation between compressions: if these simulations were too short and the density too high, then the initial FCC structure survives, and the system retains a solid-like character. While we shall discuss some results obtained from such crystal-like structures at the end of this Letter, most of the results discussed are for structures that are completely amorphous, as characterized by the pair distribution function and by the time dependent mean-squared displacement of the particles becoming diffusive.

For most of the high density simulations, we have generated three separate realizations of our polydisperse hard sphere fluid, i.e., three different distributions of particle radius, and have averaged our results over these. All our samples for densities up to $\eta = 0.59$ were run for long enough times so that they ceased to exhibit aging.[29] Typical values for $<r^2>$, at the end of our runs are at least as large as 4 even for the highest concentration reported, $\eta = 0.59$ (all the results are presented in reduced units where $\sigma$ and $\sigma/(k_BT/m)^{(1/2)}$ are the units of length and time). By this time, $<r^2>$ is found to vary linearly with time, consistent with diffusive motion. For all lower densities, $<r^2>$ was at least 10. Thus, we are confident that these simulations are equilibrated to the best of our abilities. Our simulations for $\eta = 0.60$, however, continue to show aging behavior for all the times that we could simulate. We therefore do not present results for this highest density. We conjecture that, for our simulations, the glass transition density is in the vicinity of $\eta = 0.60$, meaning only that equilibration becomes prohibitively long at this point (see below). No thermodynamic transition event is implied to occur at this density.

We first test our simulations against previously established findings as a check on our computational method. Figure 1a shows plots of the diffusion coefficient D, as a function of $\eta$ derived from the linear portion of plots of $<r^2>$ vs. time. The D values, which typically have an error of 10% at the highest $\eta$, agree within numerical uncertainty with literature values[30] for $\eta < 0.55$. The D results over the whole $\eta$ range are well described by a Vogel-Fulcher functional form with $\eta_{VFT} = 0.614$ [Figure 1a]. These results can also be fit well fit to a power law having a 'critical density', $\eta_c \sim 0.58$ and a power law exponent $\gamma = 2.2$ [Figure 1a, inset], as in other recent simulations.[31] Next, we consider the self-intermediate scattering function [inset to Figure 1b], computed at the wavevector q* corresponding to the first peak of the static structure factor at each density. As noted in previous studies, the structural relaxation is described by a pronounced two-step decay for $\eta \geq 0.57$, and a weak shoulder develops at lower concentrations in the range $0.53 \leq \eta \leq 0.57$. The structural relaxation times, $\tau$, obtained from this plot (i.e., times at which the $S(q^*,t)$ assumes a value of $1/e$) can also be fit to a power law with a $\eta_c \sim 0.58$ and $\gamma \sim 2.2$ [Fig. 1a, inset]. Note that the error bars on the power law fits are significant, and, for example, the exponent $\gamma$ for the diffusion constant can be set to 1.7 without significant loss of fit accuracy (goodness of fits, as characterized by a $r^2$ value, were always better than 0.98). Nevertheless, the results shown in Figure 1b for the product $D\tau$ suggest that the diffusion constant and the inverse of the relaxation time obey the same power law with system density for $\eta \leq 0.57$. These conclusions accord well with experimental results on polydisperse colloids.[32] In particular, our value for the scaling exponent, 2.2, is close to the experimental result, 2.4. The absolute value of D and the caging plateau value of the intermediate scattering function are not in agreement with experiment, but our simulated model (hard sphere fluid with MD) is quite different than the experimental one (colloidal hard sphere suspension with hydrodynamic interactions). We note that at higher $\eta$ our D values are also systematically smaller than those obtained from a slightly different model using Brownian Dynamic simulations.[31] Nevertheless, we are satisfied that the qualitative behavior of real hard sphere systems are recovered by our simulations, especially in light of our quantitative agreement with benchmark calculations of ref.[30].

For densities higher than $\eta = 0.57$, we see a clear breakdown of the Stokes-Einstein equation [Figure 1b] as has been observed in numerous measurements and simulations on supercooled liquids.[1, 3-11, 15, 16, 18, 19, 21-25, 27, 33-40] Also shown in this plot is the root mean squared displacement of the particles at the structural relaxation time, $\tau$. It is clear that both $\sqrt{6D\tau}$ and $\langle r^2 \rangle_\tau^{1/2}$ nearly coincide, suggesting that the first quantity provides an accurate measure of the mean displacement of the particles at the structural relaxation time. These findings are somewhat different from the results of the Brownian dynamics simulations of Voigtmann et al.,[31] who found that the product $D\tau$ increases monotonically for $\eta \geq 0.50$. It should be noted that Voigtmann et al. used a steep but continuous interaction potential rather than a hard sphere potential. So, the Voigtmann et al. results seemingly agree with earlier studies.[41, 42] of systems with continuous, although not that steep, potentials where both Molecular and Brownian Dynamics found a breakdown of the SE relation over an extended range of a control parameter (i.e. the temperature).

A number of recent computational studies of glass forming fluids have emphasized the existence of particles that are "mobile" and "immobile" relative to average motion on the time scale of the maximum of the non-Gaussian parameter where deviations from Fickian diffusion were assumed to be maximal.[5, 18, 36] These investigations have quantified the tendency of particles of excess "mobility" to become increasingly correlated in spatial position on cooling.[5, 14, 36, 43] The development of a plateau in the intermediate scattering function seems to coincide with the emergence of this dynamic heterogeneity. In our simulations, this plateau may be present for densities as low as $\eta = 0.53$, but the crossover density is hard to precisely locate. We note that the density $\eta$ where this dynamic heterogeneity sets in is close to theoretical estimates of the mode coupling density, $\eta \approx 0.52$.[44] Since this phenomenon is well documented, we focus our attention instead on a higher $\eta$ range where basic questions about the nature of molecular transport remain. Figure 2 shows distributions of particle displacement, $r^2$, away from their initial position (t = 0) for a range of dimensionless times. We choose three different densities that are in close proximity of the 'critical value' $\eta_c \approx 0.58$, defined through the fitting of the simulation data in Figure 1a (inset) to a power law form inspired by Mode Coupling theory. (Note that we ignore the well-known discrepancy between the theoretical estimates of $\eta_c$ and the simulation values.) According to conventional reasoning, deviations from MCT should arise from rare "hopping events" that allow for relaxations above $\eta \geq \eta_c$.[3] Consistent with this hypothesis, we find clear evidence of particle displacements to distances that are integral multiples of the intermolecular spacing for times comparable to $\tau$. For $\eta = 0.58$, it is apparent that the hop-like motion persists for times as long as $10\tau$, in agreement with the MD simulations of Yamomoto and Onuki for a soft-sphere glass forming liquid.[9] For $\eta=0.59$, we find that hop-like motion persists for even longer times emphasizing the increased difficulty in equilibrating the fluid within a reasonable timescale.

Clearly "hopping" events deduced from Figure 2 require something more than the dynamic heterogeneity mentioned above since these hopping features appear only at densities much higher than the onset of dynamic heterogeneity (i.e., where a plateau in the intermediate scattering function becomes evident, $\eta \sim 0.53$). On the other hand, the rare hopping particles can certainly be classified as a kind of "mobile" particles that are particularly well defined in these congested fluids. (Note that this definition has no evident relationship to the definition of mobile particles considered by Glotzer and coworkers.) We define hopping (H) particles at a given time by the requirement $r^2 \geq 0.45$, and we define the complementary particles as "sedentary" (S). The precise numerical value 0.45 corresponds to the well defined position of the first local minimum in the plots shown in Figure 2 especially at the two higher densities, a quantity that appears to be independent of $\eta$ over the narrow ranges considered. Given this definition of H and S particles we consider how these two populations contribute to the breakdown of the SE. (Note that the sedentary particles should be related to the slow particles that have been studied by Lacevic and Glotzer[18] using a four-point correlation function; these particles are slow on a time scale comparable to the structural relaxation time that is much longer than the time scale of the maximum of the non-Gaussian parameter).

An examination of the S particles at $\tau$ reveals an important effect. These particles exhibit no detectable deviation from the SE (Figure 1b). A few points need to be stressed here: (i) The S particles are not diffusive at the structural relaxation time $\tau$ of the system as a whole. Consequently, these results represent the RMS displacement of these particles at this time, rather than the product $(6D\tau)^{1/2}$ for the slow population. (ii) The S sub-population has a self-intermediate scattering function, which is different from the quantity calculated for all the particles. This difference becomes appreciable for $\eta \geq 0.57$, and becomes larger with increasing $\eta$. Correspondingly, the structural relaxation time of the S sub-population becomes larger with increasing $\eta$. The $<r^2>$ values calculated at the structural relaxation time of all the particles or of the slow population are somewhat different, this difference being as large as ~20% at the highest density simulated. So, employing the structural relaxation time of the S sub-population would change this result by 10%, which is within the uncertainties of our calculations. With these caveats, our results strongly support the notion that the S particles behave identically (on average) at all densities, and the breakdown of the SE equation may be attributed purely to the presence of the H particles.[1,3]

Previous experimental work has established the dynamical exchange of "immobile" and "mobile" populations, and found that the exchange time is comparable to $\tau$, except in the immediate vicinity of the glass transition where it can be as much as 100 times larger.[6] We define a dynamic correlation function which can quantify the persistence of the H (or S) particle states. We establish a time origin at t=0 and proceed a time corresponding to $\tau$, the structural relaxation time. Here we "mark" all the particles that execute hops. We define a function $f_i(t) = \theta[(r_i[t+\tau] - r_i[t])^2 - 0.45]$, where $\theta$ is the step function, and $f_i(t=0)$ is unity if the particle $i$ hops between 0 and $\tau$. We then select new time origins t>0, go to a time $\tau$ from this new origin, and ask which of the marked particles continue to be H particles. We ensemble average this quantity, normalize it so that it goes from 1 to 0 as a function of time (Figure 3),

$$C(t) = \frac{N\sum_i \langle f_i(t)f_i(0)\rangle - \left(\sum_i \langle f_i(0)\rangle\right)^2}{N\sum_i \langle f_i(0)\rangle - \left(\sum_i \langle f_i(0)\rangle\right)^2}$$

We find that this autocorrelation function for the persistence of the H particle state, C(t), takes progressively larger t to decay with increasing $\eta$ and we define the mean exchange time by locating the point where this function assumes a value of (1/e). In Figure 3b we plot this exchange time, and the structural relaxation time as a function of $\eta$. Over the whole range of densities it is clear that the exchange time is shorter than $\tau$, except there is a clear trend that these two curves will cross in the vicinity of $\eta$=0.59. These results are apparently consistent with recent NMR measurements that suggest that the "lifetimes" for clusters of slow particles are comparable with $\tau$ for temperatures roughly 10K above the glass transition.[6] Further, our simulations suggest that exchange times can become progressively longer than $\tau$ but only in the immediate vicinity of vitrification. Since we are unable to equilibrate our sample for $\eta$ greater than 0.60 (a density we previously identified with the glass transition),[12, 45, 46] we find results which are qualitatively in excellent agreement with experiment.

Finally, we comment briefly on situations where the initialization of the system did not yield amorphous samples, but rather yielded specimens where the spheres remained on a FCC lattice. For densities $\eta \geq 0.58$, we find that the particle 'rattle' around their lattice positions, but these systems remain caged, and hence fixed in this structure for the entire duration of our simulations. For densities lower than $\eta \approx 0.58$, on the other hand, we find that the systems relax into an amorphous configuration by allowing for the presence of hoppers (note that the polydispersity is crucial for this process; a monodisperse FCC hard sphere system with density $\eta > 0.545$ is in a stable, single phase crystalline state). For $\eta < 0.53$, the systems melt homogeneously without any H particle motion being required. To gain more insight into these results, we have calculated the RMS value of the cage size relative to the interparticle spacing. First, we used the intermediate time plateau value of $<r^2>$ to define the cage size. In a second method we froze the particle positions at randomly chosen times and computed the minimum distance a particle would have to move to collide with another particle. The mean value of the square of these distances gives another physically motivated measure of cage size. The density where we just begin to see motion in the "crystalline" solid ($\eta \sim 0.58$), and the last density where we can equilibrate an amorphous liquids ($\eta = 0.59$), both yield cage sizes of $\approx 0.14\sigma$, consistent with the standard Lindeman criterion of melting. It appears that the Lindeman criterion of melting correctly anticipates both the density at which melting and vitrification occurs in these simple liquids.[11, 47-51]

In summary, our results clearly show that an appropriately defined slow sub-population (sedentary particles) follow the Stokes-Einstein relationship even for $\eta$ beyond the fitted $\eta_c = 0.58$, where the liquid as a whole violates this relationship. This result provides conclusive evidence that the complementary "hoppers" are responsible for the breakdown of the Stokes-Einstein relationship..

The authors thank the National Science Foundation (DMR-0422062, SK and CHE-0111152, GS) for financial support of this work. SK also thanks Ken Schweizer, Alexei Sokolov, Arun Yethiraj and Mark Ediger for discussions.

**Figure Captions**

**Figure 1**: (a) Plot of the diffusion coefficient, D, plotted as a function of η (■). The dot-dashed line is a fit to the VFT equation as discussed in the text. The dashed line is the equation presented by Speedy. (inset): Plot of $D^{(1/\gamma)}$ and $\tau^{-(1/\gamma)}$ as a function of η. The lines are best fits. (b) Plots of $\sqrt{6D\tau}$ as a function of η (■). Also plotted are the $<r^2>^{1/2}$ at the structural relaxation time of the liquid (▲). Finally we plot $<r^2>^{1/2}$ for the slow population at the structural relaxation time (●). (inset) Plots of the self intermediate function as a function of reduced time. The q* value corresponded to the first peak in the static structure factor: this quantity increased approximately linearly with η. From right to left these curves correspond to η = 0.59, 0.58, 0.57, 0.56, 0.55, 0.53, and 0.50.

**Figure 2**: Distributions of particle displacement, $r^2$, as a function of $r^2$. Data are shown for three different densities and for a variety of times t = 0.1, 1, 10, 100, etc. The structural relaxation times for each density are also reported in the figure.

**Figure 3**: (a) Plots of the correlation function for exchange, C(t), as a function of t. From right to left they correspond to densities of η=0.59, 0.58, 0.57 and 0.56. (b) Plots of exchange time (denoted by the time where C(t)=1/e) and structural relaxation time as a function of η. Lines are guides to the eye.

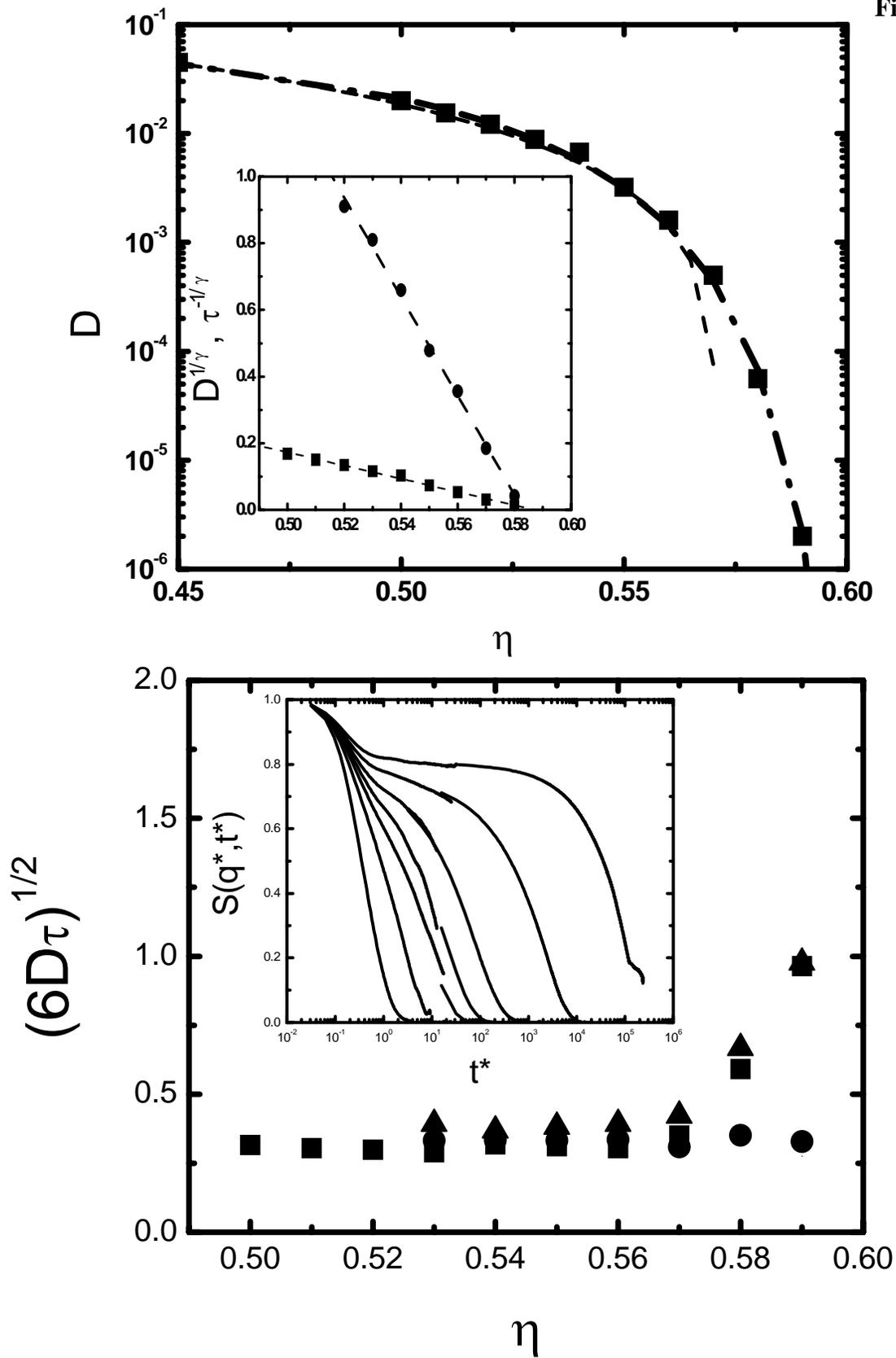

Figure 1

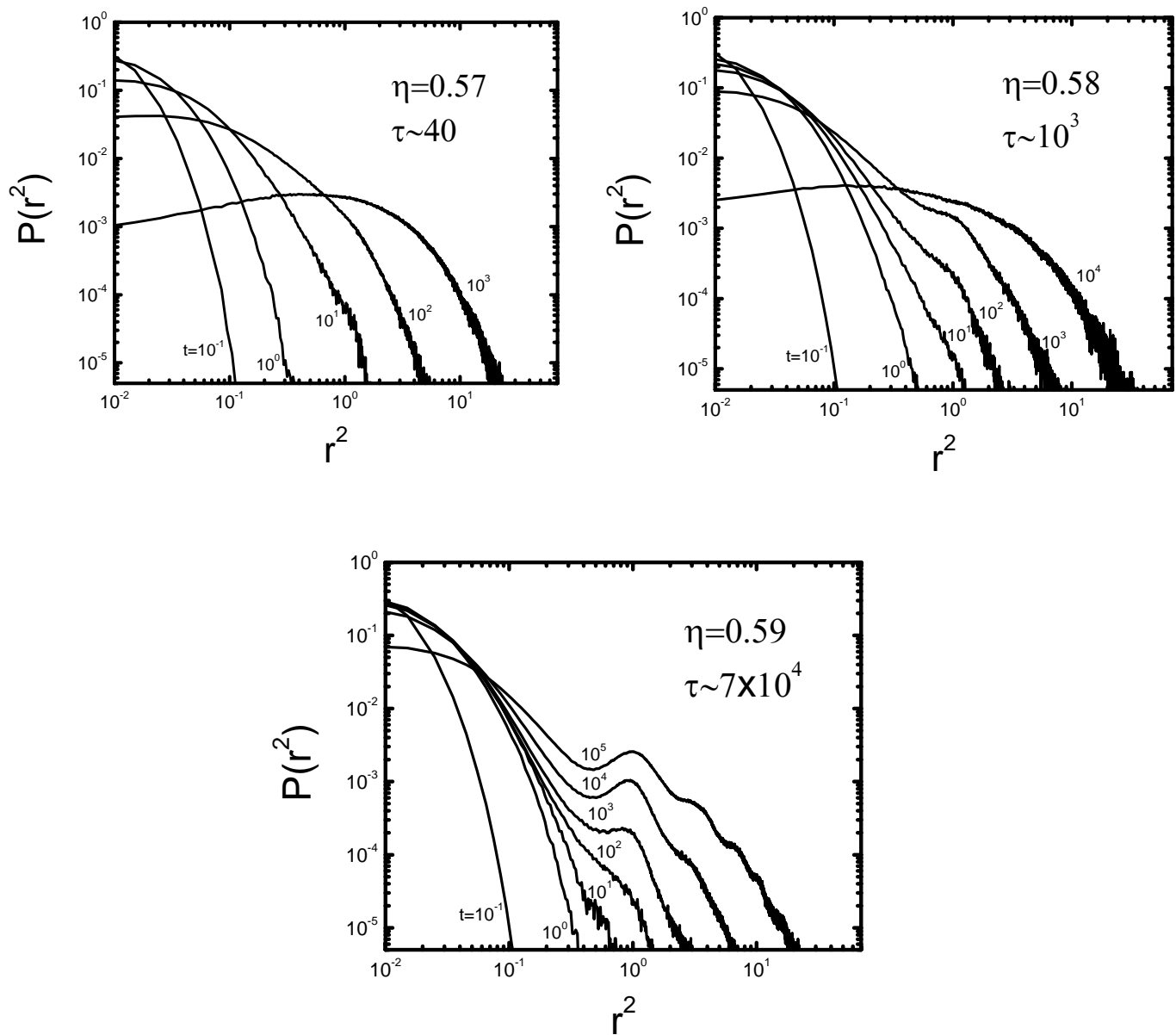

**Figure 2**

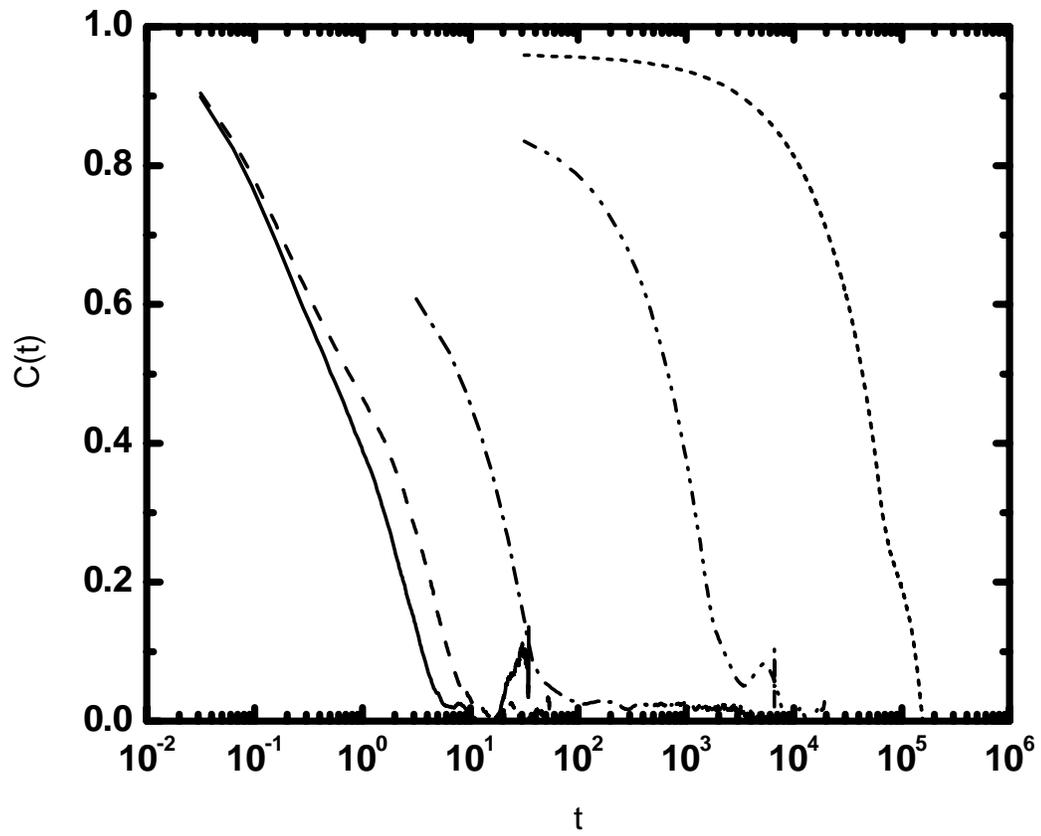

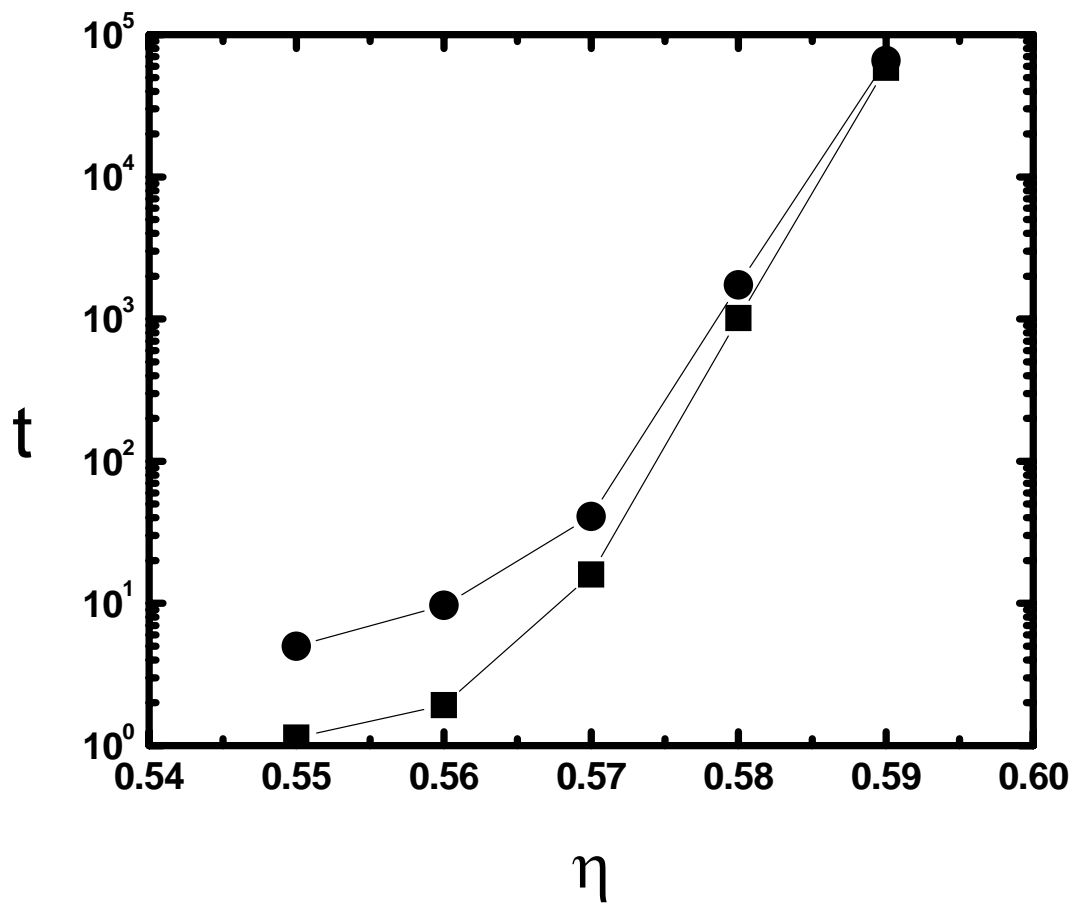